\documentclass[runningheads]{svmult}

\usepackage{makeidx}   
\usepackage{graphicx}  
\usepackage{psfig}                      
\usepackage{subeqnar}  
\usepackage{multicol}  
\usepackage{physprbb}  
\makeindex             

\def\ltsima{$\; \buildrel < \over \sim \;$}
\def\simlt{\lower.5ex\hbox{\ltsima}}
\def\gtsima{$\; \buildrel > \over \sim \;$}
\def\simgt{\lower.5ex\hbox{\gtsima}}

\begin{document}
\title*{X-rays from Star-Forming Regions
\protect\newline in the VLT Era}
\toctitle{X-rays from Star-Forming Regions in the VLT Era}
%
%
\titlerunning{X-rays from Star-Forming Regions in the VLT Era}
%
\author{Thierry Montmerle\inst{1}
\and Nicolas Grosso\inst{2}}
\authorrunning{T. Montmerle and N. Grosso}
%
%
\institute{Service d'Astrophysique, CEA Saclay, 91191 Gif-sur-Yvette, France
\and Max-Planck Institut f{\"u}r Extraterrestrische Physik, D-85741 Garching, Germany}

\maketitle              

\begin{abstract} 
The association between star-forming regions and X-ray emission
was discovered over 30 years ago.  We now know that essentially all young
stellar objects, T Tauri stars and protostars, are X-ray emitters, although the
case of the youngest, Class~0 protostars, is less clear.  The paper highlights
X-ray emission and absorption mechanisms, and summarizes X-ray observations of
young stellar objects. The impact of the VLT on the characterization of the new 
X-ray sources is also briefly discussed.  
\end{abstract}

\section{From past to present}

The first evidence that X-rays could reach the Earth from outer space, and more
specifically from the Sun, was obtained in 1949 by means of a V2 rocket launched
from White Sands, New Mexico, by a team of physicists from the US Naval Research
Laboratory; the flight lasted a mere 336\,s (Friedman, Lichtman, \& Byram 1951).
The real astronomical start took place with the launch of the {\sl Uhuru}
satellite, in 1970.  This satellite, operating for three years in scanning mode
with collimators, obtained the first X-ray all-sky survey, with the discovery of
339 sources, with error boxes up to several $10'$ on a side.  One of them
included M42, the Orion nebula (Giacconi et al.  1972):  {\it the knowledge that
star-forming regions are emitting X-rays is nearly 30 years old !  } The
discovery of the X-ray emission from M42 was confirmed by the ANS satellite (den
Boggende 1978).  The second major leap was the launch of the second
``High-Energy Astronomical Observatory'', {\sl Einstein}, in 1978, which carried
X-ray focussing mirrors able to obtain the first pointed images of astronomical
X-ray sources, with a field-of-view of $\sim 1^{\circ} \times 1^{\circ}$.  Orion
and other star-forming regions were successfully detected, with a combination of
point sources identified with OB stars and/or T Tauri stars (Ku \& Chanan 1979).
Launched in 1990, ROSAT undertook an all-sky survey, which resulted in a catalog
of several $10^5$ sources, including $\sim$50,000 stars in all evolutionary
stages, from PMS to evolved.  
Launched in 1993, the Japanese satellite ASCA was the first imaging satellite 
using CCD detectors and reaching the hard X-ray range.
To this date, four X-ray satellites are in orbit:
BeppoSAX, RossiXTE, and the two ``workhorses'' of the present decade, the US
{\sl Chandra} and the European {\sl XMM-Newton}, both launched in 1999.  These
two major observatories currently have significant programs dedicated to the
study of young clusters and star-forming regions.  Tables 1 and 2 give the
current list of accepted targets, updated from Sciortino (2001).

\newpage

\begin{table}[h!]
\begin{center}
\label{tab_sci:1}
\caption{SFRs and young stellar clusters in the {\sl Chandra} 
AO-1 Garanteed Time Observation (GTO) and Guest Observation (GO) 
AO-1, AO-2 [left] and AO-3 [right] programs (June 2001)}
\begin{tabular} {l l | l l }
\\
\hline
Target Name & Exposure Time &  Target Name & Exposure Time \\
\hline
NGC 1333 &  50\,ks               & IRAM04191     & 20\,ks \\
IC 348 &  50\,ks                 & L1527         & 20\,ks \\
Orion Trapezium &  50\,ks + 69\,ks & $\alpha$~Tau    & 20\,ks \\
Orion region & several obs      & ONC Flanking Field S & 50\,ks \\
R CrA Cloud Core & 20\,ks        & ONC Flanking Field N & 50\,ks \\
$\rho$~Oph & mosaic of 6 $\times$ 5\,ks & $\sigma$~Ori   & 100\,ks \\
$\rho$~Oph  &  2 $\times$ 100\,ks       & HH 24-26      & 70\,ks \\
NGC2516 &  20\,ks + 20\,ks           & Maddalena's Cloud & 30\,ks \\
NGC2516 &  20\,ks + 50\,ks           & NGC2264       & 50\,ks + 100\,ks \\
M16 & 80\,ks                     & NGC2362       & 47\,ks + 48\,ks \\
NGC 6530 & 61\,ks                & IRAS16293     & 30\,ks \\
W3B & 40\,ks                     & Trifid Nebula & 60\,ks \\
HH~1 & 20\,ks                    & HH~80/81      & 40\,ks \\

\hline
\end{tabular}
\end{center}
\end{table}

\vspace*{-1cm}

\begin{table}[h!]
\begin{center}
\label{tab_sci:2}
\caption{SFRs and young stellar clusters in the {\sl XMM-Newton/EPIC} 
GTO and AO-1 GO program (June 2000).}
\begin{tabular} {l l | l l}
\\
\hline
Target Name & Exposure Time & Target Name & Exposure Time \\
\hline
NGC 2024 & 45\,ks + 25\,ks        & L1448-C & 30\,ks \\ 
NGC 2023 & 25\,ks                & NGC 1333 & 50\,ks \\ 
IC 2602 & 45\,ks                 & Orion P1795 & 20\,ks \\ 
IC 2391 & 45\,ks                 & OMC2/3 & 100\,ks \\
alpha Per & 50\,ks               & L1641-N & 50\,ks \\ 
Upper Sco-Cen & 2 $\times$ 50\,ks       & NGC 2264 & 2 $\times$ 35\,ks \\ 
R CrA  & 20\,ks                  & Chamaeleon & 28\,ks \\
$\rho$~Oph & 25\,ks + 50\,ks      & Serpens & 50\,ks \\ 
IC 348 & 40\,ks                  & NGC 2362 & 50\,ks \\
L1551 & 50\,ks                   & NGC 2547 & 50\,ks \\
Blanco 1 & 50\,ks                & Pleiades & 2 $\times$ 50\,ks \\
\hline
\end{tabular}
\end{center}
\end{table}

The latest review of pre-{\sl Chandra}/{\sl XMM-Newton} 
results on Young Stellar Objects (YSOs) is
by Feigelson \& Montmerle 1999 (FM); the review by Glassgold, Feigelson \&
Montmerle (2001; hereafter GFM) puts more emphasis on the various physical processes at
work in YSOs and in their vicinity.  The reader is referred to these reviews for
details.  In the present short contribution, we only give some key background
information, summarize and update the main X-ray results, and emphasize the role
of the VLT in the problem of the characterization of the new X-ray sources.

\section{X-ray emission and absorption mechanisms}

$\bullet$ {\it YSO X-ray emission} comes from the thermal bremsstrahlung
emission of an optically thin, ``coronal'' plasma at temperatures 
$T_\mathrm{X} \sim 10^6$--$10^8$\,K and densities 
$n_\mathrm{e} \sim 10^{10}$--$10^{12}$\,cm$^{-3}$ (FM).  It can be
shown that the emitting plasma must be confined in magnetic loops, like on the
Sun, but much larger.  Stellar X-ray spectra also indicate the presence of lines
of heavy elements (see examples in, e.g., Audard et al.  2001), so that
abundances and metallicities can be determined from $\chi^2$-type fits.  This is
also the case for the X-ray brightest young stars.

\noindent$\bullet$ {\it X-ray absorption} is an important parameter in all studies
of young stars (GFM). It is due to the photoelectric effect, in 
which inner-shell (mostly K and L) electrons of an atom are ejected by 
the incoming photon.  However, these photoelectrons basically carry 
away the energy of the original X-rays (in the $\sim$ keV range), minus 
the shell energy, so are themselves initially highly energetic compared 
to the energy of the outer shells of the ambient atoms.  A ``shower'' 
results, in which many secondary electrons are ejected from ambient 
atoms as the primary electrons collide with them and loose their 
initial energy.  Other secondary electrons may be produced as ``Auger 
electrons'', i.e., from internal rearrangement of the energy levels of the 
original atoms hit by the X-ray photons.  

\begin{figure}[ht] 
\begin{center} 
\leavevmode
\psfig{file=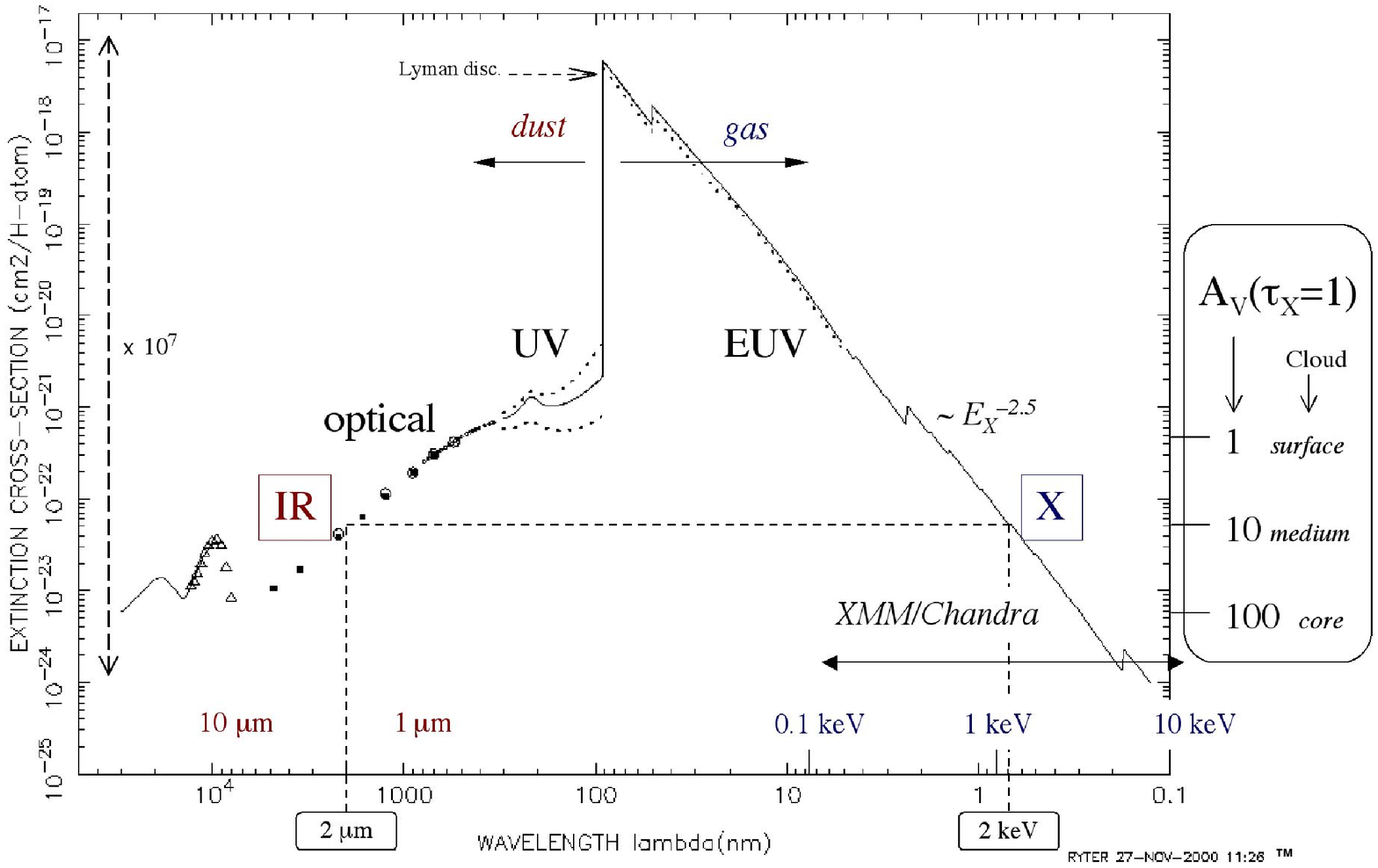,angle=0,width=\textwidth} 
\end{center} 
\caption{\small {Total extinction cross section, from the IR range (1\,$\mu$m 
$\equiv$ 1000\,nm), to the X-ray range (1\,keV $\equiv$ 0.81\,nm, using $E 
= h \times \lambda^{-1}$, with $h$ the Planck's constant) (see Ryter 1996). 
The sharp peak in the middle is the Lyman discontinuity (almost four orders of magnitude !).}} 
\label{fig:IRX} 
\end{figure}

Fig.~\ref{fig:IRX}, adapted from Ryter (1996), gives in graphic form the total 
extinction cross-section all the way from the far-IR to the hard X-ray 
range, assuming a normal gas-to-dust ratio and cosmic abundances.  
First, it can be noted that {\it the cross-section has comparable values in 
the IR range} (where it is dominated by the dust) {\it and in the X-ray range} 
(where it is dominated by the gas):  in other words, the ``penetrating 
power'' of X-rays is the same as in the IR, a key point allowing detailed 
studies of embedded sources such as young T~Tauri stars (hereafter TTS) 
and protostars (see the detailed discussion by Casanova et al.  1995).  
Second, the total photoelectric cross-section, which for each element has the form 
$\sigma \propto E_\mathrm{X}^{-2.5}$ beyond the Lyman limit (H), displays 
discontinuities.  With increasing X-ray energy, each ``jump'' corresponds 
to the inner-shell ionization of a new atomic species present in 
gaseous form in the ISM:  He, C, N, O below 1\,keV, up to Fe and Ni above 
$\sim 7$\,keV.  Although these cross-section jumps generally are small, 
they accumulate to the point that above 0.5\,keV the absorption comes 
predominantly from heavy atoms, and is $> 10$ times larger than from 
H+He beyond $\sim 7$\,keV (for details, see Wilms, Allen, and McCray 
2000).

For $E_X=1$\,keV and solar abundances, one finds $\tau(1\,{\rm keV}) = N_{\rm
H}/4.41 \times 10^{21}$\,cm$^{-2}$.  If the usual conversion between $N_\mathrm{H}$ and
$A_\mathrm{V}$ for diffuse interstellar clouds is used (e.g., Ryter 1996), then
$\tau(1\,{\rm keV}) = 1$ occurs for $A_\mathrm{V} \simeq 2$.  It is also a useful
rule-of-thumb that the penetrability of X-rays varies like $\sim E_\mathrm{X}^{-2.5}$:
the absorption decreases by a factor $\sim 3000$ between 0.4\,keV and 10\,keV.
Note that the contribution of dust grains (into which certain heavy atomic
species are condensed) to the X-ray opacity is small, and depends on grain
models only at the lowest energies ($E_\mathrm{X} \simlt 0.3$\,keV) (see discussion in
Wilms, Allen, and McCray 2000).

\section{X-ray observations of low-mass star-forming regions}

\begin{figure}[ht]
\centerline{\hbox{\psfig{file=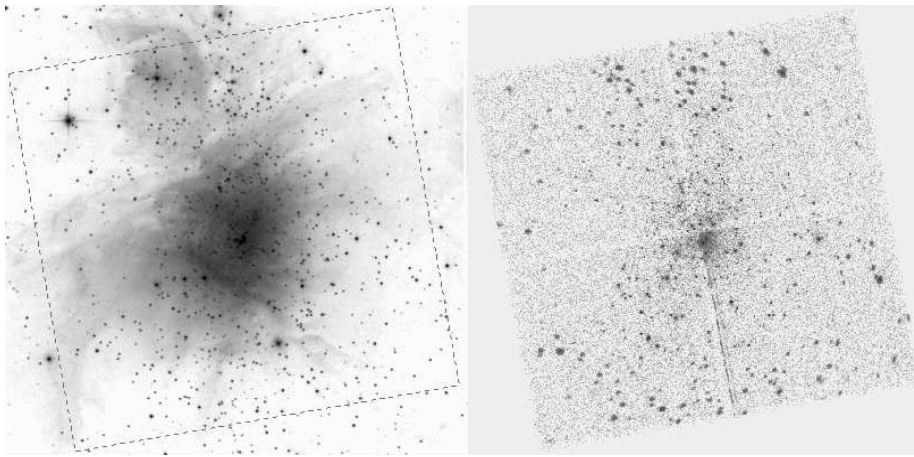,width=\textwidth}}}
\caption[]{\small {{\it Left:} The Orion Nebula Cluster seen at 2\,$\mu$m, 
from the
2MASS survey. The Trapezium O stars exciting the M42 nebula are located at
the center of the image. {\it Right:} The same region seen by {\sl Chandra/ACIS}.
Almost 1000 X-ray sources are detected (Garmire et al. 2000); 
there is nearly a one-to-one subarcsecond
correspondence between the near-IR and X-ray sources in both images.
Field of view of {\sl Chandra/ACIS}: $17' \times 17'$.}}
\label{fig:Orion-IRX}
\end{figure}

To date, all major star-forming regions within $\sim 500$\,pc of the Sun have
been observed in X-rays, plus a couple of more distant ones ($d \geq 1$\,kpc)
(see FM).  Perhaps the most spectacular observation to date is that of the Orion
Trapezium region and its surrounding M42 nebula (Garmire et al. 2000), in which
a single $17' \times 17'$ {\sl Chandra} field has revealed nearly 1,000 sources.
As shown on Fig.~\ref{fig:Orion-IRX}, almost every X-ray source has a stellar
counterpart (position accuracy $\sim 0.1''$), all the way from the central
massive stars down to lower-mass stars.  A full analysis is in progress
(Feigelson et al., in preparation).

\begin{figure}[ht]
\centerline{\hbox{\psfig{file=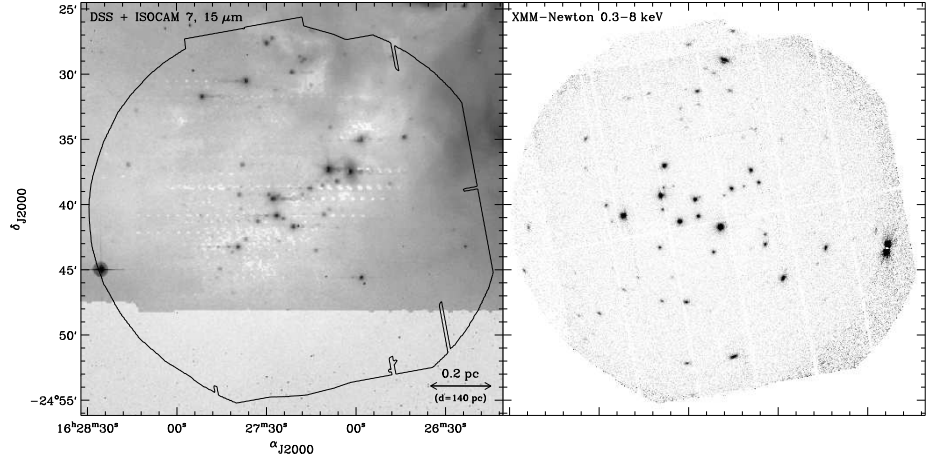,angle=90,width=\textwidth}}}
\caption[]{{\it Left:} Field centered on the core F region of the $\rho$~Ophiuchi dark cloud.
Background: DSS optical image. Foreground: ISOCAM 7+15\,$\mu$m image
(Abergel et al. 1996). Quasi-circular contour: {\sl XMM-Newton/EPIC} field-of-view.
Most of the bright ISOCAM sources are protostars, the other sources
are T Tauri stars (see Bontemps et al.\ 2001). 
{\it Right:} {\sl XMM-Newton} image of the same region. With a few exceptions, almost
all the IR sources, including protostars, are detected (Grosso et al., 
in preparation).}
\label{fig:Oph_XMM}
\end{figure}

Here we restrict our discussion to the numerous X-ray detected low-mass,
solar-like stars.  The Taurus-Auriga, R Coronae Australis, Chamaeleon, and
$\rho$ Ophiuchi regions have been particularly well studied for many years.  For
example, ROSAT observations of the central region of the $\rho$ Ophiuchi dark cloud have
revealed several dozen X-ray sources associated with young stellar objects
(Grosso et al. 2000).  This region has now been also observed in X-rays by {\sl
Chandra} (Imanishi et al.  2001), and by {\sl XMM-Newton} (Grosso et al., in
preparation:  see Fig.~\ref{fig:Oph_XMM}).  ROSAT being sensitive in the soft
X-ray band only ($< 2.4$\,keV), the sources are mainly identified with
objects having low to intermediate extinction (see Fig.~\ref{fig:IRX}), i.e., TTS.
Nevertheless thanks to X-ray flares, which combine an increase of 
X-ray luminosity with an increase of plasma temperature hence a better visibility, 
ROSAT caught X-rays from two protostars (see Grosso et al.\ 1997 and Grosso 2001).
In contrast ASCA was sensitive up to 10\,keV, hence was able to penetrate deep 
into dense material, revealing more easily X-rays from protostars.
The past generation of X-ray satellites detected protostars only during their 
high activity level.
Today, the new generation of X-ray satellites, {\sl Chandra} and {\sl XMM-Newton}, 
with an increased sensitivity up to 10\,keV, gives access to the quiet coronae 
of protostars. 

Let us now summarize and update briefly some key X-ray properties of 
TTS and protostars.

$\bullet$ {\it There is a strong correlation between the X-ray and bolometric
luminosities of TTS,} which differs somewhat from region to region:
$L_\mathrm{X}/L_\mathrm{bol} \sim 10^{-4}$, as compared with $\sim 10^{-6}$ for the active Sun.
Typical TTS X-ray luminosities are $L_\mathrm{X} \sim 10^{30}$--$10^{31}$\,erg\,s$^{-1}$
(e.g., Grosso et al.  2000).  The presence of disks does not seem to influence
the TTS X-ray properties, but as TTS lose their disks while contracting towards
the main sequence, they become more X-ray luminous, presumably because of the
increased rotation velocity and resulting enhanced dynamo (Stelzer \&
Neuh{\"a}user 2001).  X-ray surveys of star-forming area are very efficient to
make a reliable census of the TTS population.  Nevertheless IR follow-ups are
needed to confirm the nature of the X-ray selected objects, and their
young age.

$\bullet$ {\it TTS X-ray variability is generalized.}  Peak X-ray luminosities
may be as high as $L_\mathrm{X,peak} \sim 10^{32}$\,erg\,s$^{-1}$ or more.  The light
curves are strongly suggestive of flares.  In the best cases, flare spectroscopy
can be time-resolved, showing a clear decline in temperature (from $T_\mathrm{X}$ as high
as $\sim 10^8$\,K down to a few $10^7$\,K~:  see the example of V773 Tau,
Tsuboi et al.  1998).

$\bullet$ {\it X-ray detections have reached the brown dwarf limit.}  
A few {\it bona fide} brown dwarfs have been detected in X-rays, already by ROSAT
(Neuh{\"a}user et al. 1999), and recently by {\sl Chandra} (Rutledge et al.\ 2000; 
Imanishi et al.  2001) and {\sl XMM-Newton} (Grosso et al., in preparation).  
Their $L_\mathrm{X}/L_\mathrm{bol}$ ratio is found to be $\simgt 10^{-4}$, i.e.,
very similar to that of TTS.  It is reasonable to think that a number of new
brown dwarf candidates will soon be {\it X-ray detected} among {\sl Chandra} 
and {\sl XMM-Newton} sources in star-forming regions.
An IR spectroscopy confirmation will be necessary, and will require large 
telescopes.


$\bullet$ {\it Deeply embedded sources like protostars are more difficult to
detect in X-rays.}  The number of protostars detected remained small (about a
dozen) in the ROSAT/ASCA era, for lack of access to the hard X-ray range (ROSAT), 
or limited sensitivity (ASCA) (FM).
The new {\sl Chandra} and {\sl XMM-Newton} results on the $\rho$ Ophiuchi dark cloud (see
above), are very promising, with a $\sim$ 70 $\%$ detection rate, including at
their non-flare, quiescent level.

$\bullet$ {\it Almost all of the $\sim 15$ {\rm {bona fide}} protostars detected
so far in X-rays are evolved (``Class I'') protostars.}  Although magnetism also
plays a central role in their X-ray emission, significant differences exist with
that of TTS~:  most detected Class I protostars have X-ray luminosities and
temperatures somewhat higher than those of most TTS.  Variability is as
ubuquitous as for TTS, with many examples of flares.  In a few cases
$L_\mathrm{X,peak}$ reaches very high values of several $10^{32}$\,erg\,s$^{-1}$ or more,
which may be due to star-disk magnetic interactions (Tsuboi et al.  2000,
Montmerle et al.  2000).  In contrast, the X-ray detection of young (``Class
0'') protostars is open to discussion.  Tsuboi et al.  (2001) recently reported
the {\sl Chandra} discovery of two weak X-ray sources in the direction of
candidate Class 0 protostars in the Orion OMC 2/3 clouds, but their extinction, measured
from the X-ray spectrum itself, is nearly one order of magnitude smaller than
the typical extinction towards the central object of a Class 0 source across its
own envelope.  They may therefore be somehow associated with the Class 0
protostars, but not be strictly identified with them.

\section{The role of the VLT}

Thanks to their much improved sensitivity and resulting improved performance in
spectral and spatial resolution, {\sl Chandra} and {\sl XMM-Newton} allow
detailed studies of previously known, X-ray bright YSOs like TTS and
protostars.  But a new frontier appears:  {\it low-luminosity objects}.  With a
few exceptions, the $L_\mathrm{X}/L_\mathrm{bol} \sim 10^{-4}$ correlation approximately holds
empirically for all YSOs, so that in general weak X-ray sources (now detected at
the level of $L_\mathrm{X} \sim 10^{28}$\,erg\,s$^{-1}$, i.e., only 10 times the active
Sun, at $d \sim 150$\,pc) are expected to be intrinsically low-luminosity
sources, i.e., of low mass and/or low temperature.  Very Low-Mass objects
and brown dwarfs are typical examples.  

However, contrary to the situation holding with the past generation of X-ray
satellites, {\it low-luminosity sources are not necessarily YSOs !}  The reason
is that the sensitivity of {\sl Chandra} and {\sl XMM-Newton} is such that weak,
hard X-ray sources like AGNs start to be visible through molecular clouds.  From
the extragalactic log$N$-log$S$ curve of the Hubble Deep Field seen by {\sl
Chandra}, the contamination is estimated to be $\approx 20$ background sources
in the $17' \times 17'$ field of view (Garmire et al.  2001).  Of course, such
sources must be among the weakest and most absorbed of the sample, mimicking
very low-mass stars.  Note that the same situation holds in the IR, where
background stars and galaxies routinely contaminate the young star sample,
except in the densest regions.

It is therefore clear that one needs {\it both} X-ray and IR observations to
disentangle the source sample, and large ground-based telescopes are mandatory
in almost all cases.  More generally, we have also seen (\S 2) the deep
connection existing between the X-ray and IR ranges.  Therefore, the VLT, with
its large choice of IR cameras, is certainly the facility of choice for
follow-ups of the sensitive X-ray observations in progress.

A foretaste of the discoveries to come is offered by the preliminary results of
an NTT/SOFI campaign to search for counterparts of {\sl XMM-Newton} new X-ray sources
without 2MASS counterpart in the $\rho$~Ophiuchi dark cloud (see Fig.~\ref{fig:Oph_XMM}; 
Grosso et al., in preparation).  In the course of this study, two spectacular 
new embedded Herbig-Haro objects were found, one probably related to the peculiar X-ray
emitting protostar YLW15, the other to the weaker X-ray emitting protostar, IRS54
(Grosso et al.  2001).  The campaign has also yielded very low-luminosity
IR counterparts for faint {\sl XMM-Newton} sources, the nature of which can be
elucidated only with follow-up IR spectroscopy such as provided by
ISAAC in the near-IR or VISIR in the mid-IR.  Could these be ``free-floating
planets'' so well appreciated by Mark McCaughrean ?  (After all, Jupiter is also
an X-ray source, although of auroral origin...)

%

\end{document}